\documentclass[]{ceurart}

\usepackage{tikz}
\usetikzlibrary{arrows.meta, positioning, calc, fit, backgrounds, decorations.pathmorphing}
\usepackage{xcolor}

\definecolor{capblue}{HTML}{3B7DD8}
\definecolor{recgreen}{HTML}{27AE60}
\definecolor{revizorange}{HTML}{E67E22}
\definecolor{feedbackred}{HTML}{C0392B}

\begin{document}

\copyrightyear{2026}
\copyrightclause{Copyright for this paper by its authors.
  Use permitted under Creative Commons License Attribution 4.0
  International (CC BY 4.0).}

\conference{SAXR '26: Shaping Future Human Connection: Social Augmentation through XR Technologies – 1st Workshop edition at CHI Conference on Human Factors in Computing Systems,
  April 2026, Barcelona, Spain}

\title{The Attention-Aware Pipeline: Design Tensions from Making Attention Visible in XR}
\shorttitle{The Attention-Aware Pipeline}
\shortauthors{Srinivasan and Elmqvist}

\author[1]{Arvind Srinivasan}[%
orcid=0009-0007-0565-7054,
email=arvind@cs.au.dk,
]
\author[1]{Niklas Elmqvist}[%
orcid=0000-0002-3985-4990,
email=elm@cs.au.dk,
]

\address[1]{Department of Computer Science, Aarhus University, Denmark}

\begin{abstract}
    Where people look during shared activity carries coordination cues that speech and gesture cannot replace, but these patterns remain invisible to participants.
    XR headsets make gaze available as real-time input, yet few systems feed it back visually.
    We frame our work using the Attention-Aware Pipeline (Capture, Record, Revisualize), whose feedback loop means the system's visual response alters what users attend to next, triggering further responses.
    This generates design tensions whose form depends on each stage's configuration.
    We trace the pipeline through three systems casting attention as a \emph{mirror} (reflecting gaze history), a \emph{medium} (sharing it across collaborators), and a \emph{mediator} (intervening through diminished reality).
    Each encountered a tension the loop predicted, motivating the next.
    A formative eye-tracking study of four musicians surfaced attentional tunneling and near-total disconnection, confirming the need for intervention.
    We present these tensions and a next step: testing whether subtractive intervention reduces tunneling for a single sight-reader.
\end{abstract}


\maketitle

\section{Introduction}
\label{sec:intro}

A four-person band is rehearsing.
The bassist has been staring at the lead guitarist for ten minutes straight, with 66\% of all visual attention locked on a single person.
The vocalist has not looked at the bassist once.
The drummer scans everyone roughly equally.
The guitarist, aware that the bassist needs help to stay on-time with the track, reciprocates with 50\% attention to count in.
None of them knew any of this was happening until after.
These are real numbers from a formative needfinding study we describe in Section~\ref{sec:needfinding}.

If you could make these patterns longitudinally visible to the individuals, you would have a powerful coordination tool.
The bassist could see that their attention has narrowed.
The vocalist could discover they have been ignoring a collaborator.
The guitarist could see the asymmetric coordination burden they carry and respond appropriately.
But \emph{how} you make attention visible matters, because doing it naively can make things worse.
Our own research shows that adding visual overlays to convey attention helps in sparse settings but creates interference in dense ones---the more complex the scene, the more coordination support is needed, yet the less room there is for additive cues.

By \emph{attention}, we mean the selective concentration of perceptual and cognitive resources on a subset of available stimuli~\cite{DBLP:journals/tvcg/SrinivasanEBRE25}.
In visual tasks, attention is most directly indexed by gaze: where a person fixates, for how long, and in what sequence reveals what they are prioritizing and what they are missing.
Eye-tracking hardware in current XR headsets makes this signal continuously available, turning gaze from a diagnostic measure into a real-time input channel.
An \emph{attention-aware} system~\cite{DBLP:journals/tvcg/SrinivasanEBRE25} is one that captures this gaze signal, records it over time, and revisualizes the result back into the user's environment.
This closes a feedback loop: the system's visual response changes what the user attends to, which changes what the system captures next, which changes the next response.
We organize the design space of such systems using a 4W1H questioning framework (\emph{Who, What, Where, When, How}) applied to each stage of the pipeline, which surfaces the specific design decisions and tensions that arise at each step (Section~\ref{sec:pipeline}).

We have been working on this problem across three projects, each of which encountered a design tension that the previous system's configuration could not resolve.
Attention-Aware Visualizations (AAVs)~\cite{DBLP:journals/tvcg/SrinivasanEBRE25} treat attention as a \emph{mirror}: the system reflects a user's gaze history back into the visualization.
HeedVision~\cite{Srinivasan2025heedvision} treats attention as a \emph{medium}: shared attention maps let collaborators see where others have looked.
Our current work treats attention as a \emph{mediator}: instead of adding visual cues, the system intervenes on dysfunctional gaze patterns by \emph{removing or attenuating} visual content in the scene, an approach drawn from diminished reality~\cite{Mori2017}, which modifies a user's perception by subtracting elements from the environment rather than augmenting it.

Each project encountered a design tension that motivated the next.
This paper explains \emph{why} those tensions arose and what they mean for systems not yet built.
The mechanism is a feedback loop in what we call the \emph{Attention-Aware Pipeline}:
Capture $\to$ Record $\to$ Revisualize, where the output of Revisualize becomes the input to Capture (Section~\ref{sec:pipeline}).
We trace how each of our systems maps onto this pipeline and what tension each encountered (Section~\ref{sec:configs}).
A formative needfinding study with four musicians grounds the trajectory in real attentional data (Section~\ref{sec:needfinding}) while also teaching us that the group problem is too complex to tackle directly, motivating a scoped next step: testing whether subtractive intervention can reduce individual tunneling during sight-reading (Section~\ref{sec:individual}).
We close with points of discussion for the workshop.


\section{The Attention-Aware Pipeline}
\label{sec:pipeline}

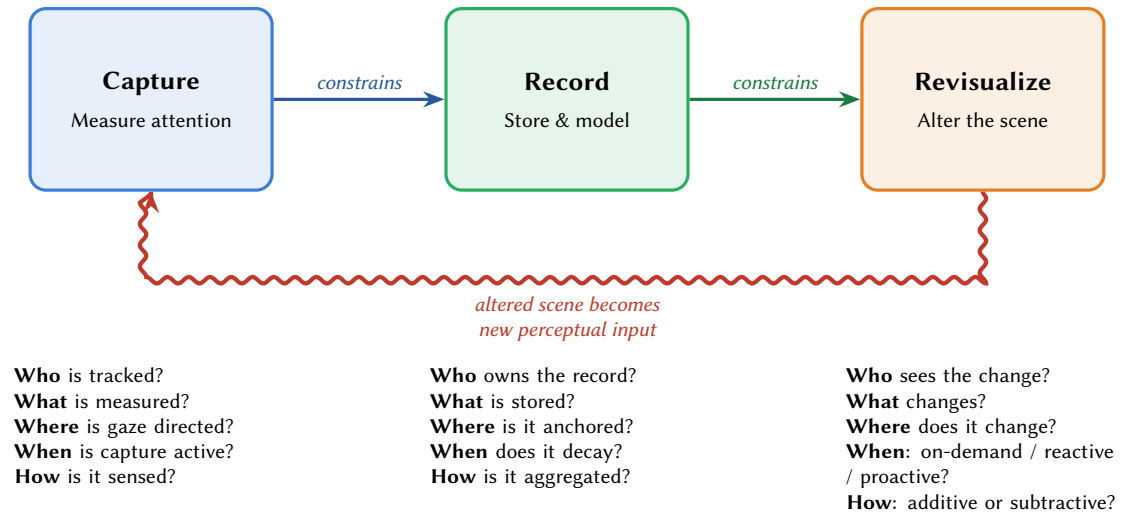
\begin{figure*}[t]
    \centering
    \begin{tikzpicture}[
        stage/.style={draw, rounded corners=6pt, minimum width=3.2cm, minimum height=2.4cm, align=center, line width=1.2pt, font=\small},
        arrow/.style={-{Stealth[length=8pt]}, line width=1.2pt},
        qlabel/.style={font=\scriptsize, align=left, text width=3.6cm},
        feedlabel/.style={font=\scriptsize\itshape, text=feedbackred, align=center},
    ]
    
    \node[stage, fill=capblue!12, draw=capblue] (cap) at (0,0) {\textbf{Capture}\\[2pt]\scriptsize Measure attention};
    \node[stage, fill=recgreen!12, draw=recgreen] (rec) at (5.5,0) {\textbf{Record}\\[2pt]\scriptsize Store \& model};
    \node[stage, fill=revizorange!12, draw=revizorange] (rev) at (11,0) {\textbf{Revisualize}\\[2pt]\scriptsize Alter the scene};
    
    \draw[arrow, capblue!70!black] (cap) -- node[above, font=\scriptsize\itshape] {constrains} (rec);
    \draw[arrow, recgreen!70!black] (rec) -- node[above, font=\scriptsize\itshape] {constrains} (rev);
    
    \draw[arrow, feedbackred, line width=1.6pt, decorate, decoration={snake, amplitude=1.2pt, segment length=8pt}] 
        (rev.south) -- ++(0,-1.2) -| node[feedlabel, pos=0.25, below=1pt] {altered scene becomes\\new perceptual input} (cap.south);
    
    \node[qlabel, anchor=north] at ($(cap.south)+(0,-2.2)$) {
        \textbf{Who} is tracked?\\
        \textbf{What} is measured?\\
        \textbf{Where} is gaze directed?\\
        \textbf{When} is capture active?\\
        \textbf{How} is it sensed?
    };
    
    \node[qlabel, anchor=north] at ($(rec.south)+(0,-2.2)$) {
        \textbf{Who} owns the record?\\
        \textbf{What} is stored?\\
        \textbf{Where} is it anchored?\\
        \textbf{When} does it decay?\\
        \textbf{How} is it aggregated?
    };
    
    \node[qlabel, anchor=north] at ($(rev.south)+(0,-2.2)$) {
        \textbf{Who} sees the change?\\
        \textbf{What} changes?\\
        \textbf{Where} does it change?\\
        \textbf{When}: on-demand / reactive / proactive?\\
        \textbf{How}: additive or subtractive?
    };
    
    \end{tikzpicture}
    \caption{\textit{The Attention-Aware Pipeline.}
    Three stages process gaze data into visual interventions.
    The feedback arrow (bottom) is the source of design tensions: whatever Revisualize produces enters Capture as new perceptual content.
    Design questions at each stage (adapted from the 4W1H approach similar to prior works~\cite{Shin_TVCG_2023}) define the configuration space.}
    \label{fig:pipeline}
\end{figure*}

An attention-aware system~\cite{DBLP:journals/tvcg/SrinivasanEBRE25} that tracks attention and feeds it back into the user's visual environment can be described in three stages.
We structure these as a pipeline, applying the 4W1H questioning framework to each stage (Figure~\ref{fig:pipeline}).
This framing provides systematic coverage of the design space by guiding designers to consider \emph{What} (the visualization technique), \emph{Who} (viewer considerations), \emph{When} (timing of the intervention), and \emph{Where} (target anchoring within the visual environment) alongside the \emph{How}.

\paragraph{Stage 1: Capture.}

The system measures attention.
The questions are \emph{Who} is tracked (self, collaborators, everyone), \emph{What} is measured (fixation location, duration, saccadic patterns, blink rate as a cognitive load proxy), and \emph{Where} attention is directed (at data marks, at people, at the environment). 
The \emph{Where} question is where the pipeline's assumptions break down most dramatically when moving from analytics to social settings. In AAV, attention targets are stationary data marks with clear boundaries and stable semantics.
In band practice, attention targets are moving human bodies whose boundaries are fuzzy, whose semantic content changes on a timescale of seconds (the leader's hand means something different during a chord change than during a rest), and whose relevance depends on the musical context. A bassist glancing at the leader during a tempo change is doing something functionally different from the same bassist glancing at the leader during a sustained passage, even though the gaze coordinates are identical. The system must distinguish not just \emph{where} attention is directed but \emph{what that direction means given what is happening musically}, a question that does not arise when targets are static data marks with fixed semantics. These differences propagate through the rest of the pipeline: Record must anchor attention to targets that move between frames, and Revisualize must alter the appearance of people rather than abstract marks.

\paragraph{Stage 2: Record.}

Captured attention must be stored and modeled.
\emph{Who} owns the record determines privacy: individual records stay private; shared records enable coordination but introduce surveillance concerns~\cite{Lebeck2018}.
\emph{Where} the record is anchored matters when targets move: in AAV, attention is anchored to data marks; in HeedVision, to voxel coordinates; in band practice, it would need to follow moving people.
\emph{When} the record decays controls responsiveness versus memory.
The aggregation method also matters because the same metric can mean different things depending on role.
Our formative needfinding data shows that the same dwell-time figure on the leader has different interpretations by musician.
The vocalist's 50\% may be strategic monitoring; the bassist's 66\% with elevated blink rate may signal stress-driven tunneling~\cite{Easterbrook1959}.
Role-aware recording is necessary to avoid conflating these.

\paragraph{Stage 3: Revisualize.}

The system alters the visual environment.
Three questions matter most. \emph{How}: additive (highlight, heatmap, gaze ray) or subtractive (diminish, blur, desaturate)?
\emph{Where}: at attended regions or at neglected regions?
And \emph{When}: on-demand (user requests), reactive (threshold crossing), or proactive (system predicts a problem)?

\paragraph{The Feedback Arrow.}

The feedback loop is what makes the pipeline a predictive tool rather than a post-hoc description: by tracing what happens when Stage~3's output re-enters Stage~1, designers can anticipate failure modes before building.
Whatever Revisualize produces (a heatmap, a faded mark, a dimmed region) becomes part of the visual scene.
That scene is what Capture measures next.
So the output of the pipeline is also its input.
Any closed loop creates the possibility that the system's visual response invalidates the conditions that triggered it.
The specific tension depends on which answers were chosen at each stage.
We now show that each of our three systems encountered exactly this kind of tension.

\section{Three Configurations and Three Tensions}
\label{sec:configs}

Each stage of the pipeline admits multiple configurations, and the feedback loop means that design choices at one stage propagate consequences to the others. We now trace three systems (two built, one proposed) through the pipeline.
We summarize each system's configuration as a tuple of answers to the pipeline's design questions, following the order introduced in Figure~\ref{fig:pipeline}.
Each instantiates a different metaphor for what attention-awareness is \emph{for}, and each encountered (or is predicted to encounter) a tension that the feedback loop generates.

\subsection{Mirror: Attention-Aware Visualization}

\paragraph{Configuration.}

Capture (self, fixation duration on marks) $\to$ Record (individual, anchored to data marks, temporal decay) $\to$ Revisualize (to self, at attended regions, emphasis/de-emphasis, on-demand and reactive).

\paragraph{What We Built.}

AAVs~\cite{DBLP:journals/tvcg/SrinivasanEBRE25} make a visualization responsive to what the user has perceived.
Marks already examined desaturate; unseen marks stay prominent.
The concept draws on read wear~\cite{Hill1992}, visual saliency~\cite{matzen18visualsaliency}, and adaptive visualization~\cite{Steichen13}.
In a qualitative study with 12 participants across 2D and 3D implementations, participants reported more systematic exploration and appreciated the feedback. 
The 2D version used a Pupil Labs Neon eye tracker; the 3D version used head-direction tracking on a Meta Quest headset with GPU stencil-buffer visibility detection.

\paragraph{Tension 1: The Mirror also Steers.}

De-emphasis of seen marks increases the relative salience of unseen marks.
Those attract fixation.
They get de-emphasized.
The next unseen marks become salient.
Tracing the feedback loop: the system was designed to \emph{reflect} attention, but the loop means it also \emph{redirects} it.
A pure mirror would leave perception unchanged; this mirror reshapes the salience landscape with every update.
Whether this redirection is beneficial (guiding exploration) or harmful (creating artificial fixation sequences) depends on whether the steered pattern is better than the natural one, a question the system itself cannot answer.
Once the loop is running, a system designed as a mirror begins to exhibit mediator-like behavior.

\subsection{Medium: HeedVision (Collaborative AAV)}

\paragraph{Configuration.}

Capture (group, head-directed gaze) $\to$ Record (shared voxel grid, pooled with per-user color coding) $\to$ Revisualize (to group, additive heatmap overlay, at attended regions, reactive).

\paragraph{What We Built.}

HeedVision~\cite{Srinivasan2025heedvision} extends AAV to pairs of co-located analysts in mixed reality.
We evaluated it with 16 participants (8 pairs) on Meta Quest 3 headsets, performing collaborative visual search across sparse (3D scatterplot) and dense (terrain) conditions.
In the sparse scatterplot, shared attention cues improved coordination: pairs naturally divided the search space, reduced redundant examination, and found targets more efficiently.
Participants described the effect as implicit territory formation, resembling stigmergic coordination~\cite{Grasse1959}, where agents coordinate indirectly through environmental traces rather than explicit communication.

\paragraph{Tension 2: The Attention Paradox.}

Tracing the feedback loop: the heatmap overlay is now visual content in the scene, entering Capture as a fixation target.
In the sparse scatterplot, this is helpful; the overlay is the most informative element in the scene.
In the dense terrain, the overlay competes with task-relevant visual detail.
Our gradient-attention analysis showed that terrain features already functioned as natural landmarks; the overlay added a redundant information channel that created interference.
The more complex the scene, the more you need coordination support, but the less room you have for additive cues.
This is a structural tension:
any attention-aware system that adds visual content at the Revisualize stage injects new material into Capture, and whether that material helps or hurts depends on what was already there.
The denser the scene, the greater the need for coordination support, yet the less room there is for additive cues.

\subsection{Mediator: Attention Guidance (Proposed)}

\paragraph{Configuration.}

Capture (self + context, fixation duration + blink rate, tunneling detection) $\to$ Record (individual, threshold-based episodes, anchored to dynamic targets) $\to$ Revisualize (to self, at attended region, subtractive diminishment, reactive).

\paragraph{Motivation.}

The attention paradox rules out additive overlays in visually dense environments.
But dense, dynamic environments (band rehearsals, surgical theaters, emergency coordination) are exactly where attention-based coordination matters most.
This motivated us to explore \emph{subtractive} revisualization: instead of adding cues to show where people have looked, diminishing visual content where someone has looked too long.
Diminished reality~\cite{Mori2017, Herling2010} provides the technical substrate; attention data provides the trigger.

\paragraph{Tension 3 (Predicted): The Tunneling-Target Paradox.}

Before building anything, we can trace the feedback loop and predict what could go wrong.
Suppose we diminish the tunneling target (the leader, from the bassist's perspective).
The bassist's gaze shifts.
To where?
Ideally, to other band members being ignored.
But possibly to the next-most-salient non-social target: their own hands, the music stand, or the floor.
If that new fixation triggers the tunneling threshold, it gets diminished too.
The system progressively strips the visual field.
Worse, when the tunneling target is also the coordination source, subtractive intervention is paradoxical: the bassist \emph{needs} to see the leader, just not \emph{only} the leader.
If we apply a decay mechanism, that would simply shift it back to Tension 1.

This predicted tension has a structural implication.
The group-level problem (redistribute the bassist's attention across band members) compounds the individual-level problem (reduce fixation on a single target). 
You cannot design a group-level mediator without first understanding whether subtractive intervention works at the individual level at all.
Our formative needfinding study confirms the group-level patterns that motivate this work; we then explain why the next step must focus on individual tunneling.

\section{Formative Needfinding Study: Attention in Band Practice}\label{sec:needfinding}

Our prior work evaluated attention-aware visualization in controlled analytics tasks: scatterplots, terrain maps, abstract datasets.
To test whether the pipeline's tensions hold in ecologically valid settings, we looked for a domain satisfying three criteria: continuous visual coordination between people, observable consequences when attention fails, and difficulty verbalizing gaze behavior.
Collaborative music performance satisfies all three criteria: musicians coordinate timing, dynamics, and expression largely through visual cues, yet rarely discuss their gaze behavior.
We conducted an initial, eye-tracking-based needfinding study of a four-piece band during rehearsal to characterize the attention landscape that any intervention would need to operate in.

\begin{figure}
    \centering
    \includegraphics[width=0.75\linewidth]{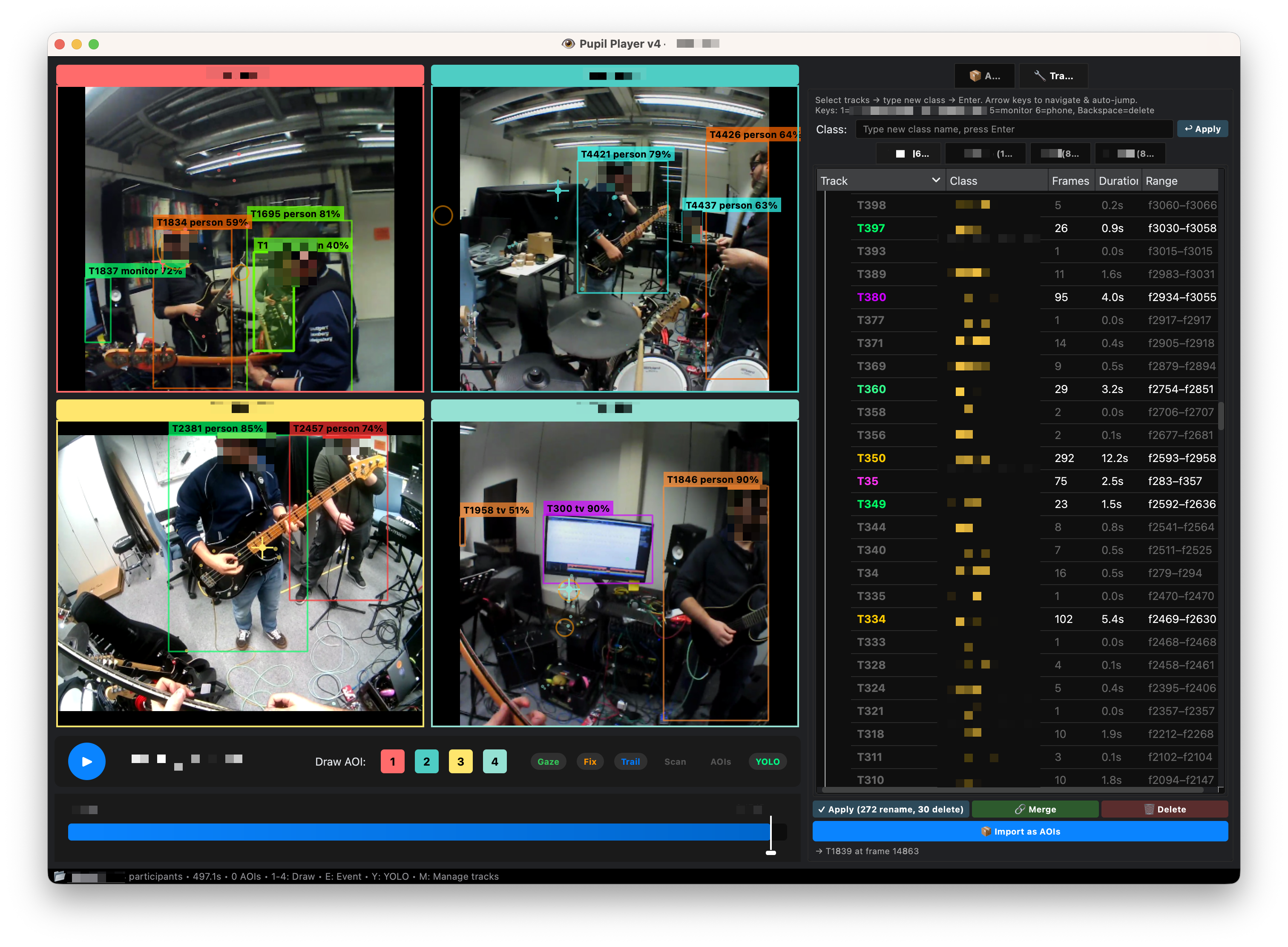}
    \caption{\textit{PupilLabs Analysis Workflow.}
    Multi-view snapshot from the band rehearsal showing dynamically tracked areas of interest (AOIs) across four musicians.
    Bounding boxes are generated via automated detection and manually refined to support gaze-to-target mapping.
    The interface illustrates the annotation workflow used to construct dynamic, person-centered AOIs required for analyzing attention directed at moving social targets.}
    \label{fig:needfinding_snapshot}
\end{figure}

\paragraph{Participants and Setup.}

Four musicians, a drummer (P01), vocalist (P02), lead guitarist/band leader (P03), and bassist (P04), rehearsed together for approximately 15--17 minutes. 
Each wore a Pupil Labs Neon eye tracker.
We defined dynamic, keyframe-tracked areas of interest (AOIs) pre-annotated by YOLOv8~\cite{yoloV8} and manually cleaned for each musician, and a shared music monitor using a custom player interface (Figure~\ref{fig:needfinding_snapshot}).
We extracted fixation sequences, dwell times per AOI, saccade characteristics, blink rates, gaze spatial dispersion, gaze transition entropy~\cite{Sato2024}, and head movement from IMU data.

We operationalized attentional tunneling through converging evidence, following Easterbrook's~\cite{Easterbrook1959} cue-utilization hypothesis and Dirkin's~\cite{Dirkin1983} cognitive tunneling framework: reduced gaze dispersion, reduced saccade amplitude, elevated dwell on a single AOI, and elevated blink rate as a physiological stress marker~\cite{Paprocki2017}.

\paragraph{Finding 1: Role-Dependent Attention Signatures.}

Attention distribution varied sharply by musical role. P01 (drummer) showed the most balanced pattern: no single target exceeded 41\% dwell time, and transition entropy was highest among the four players at 3.86 bits, indicating the most varied scanning across all band members.
P03 (leader) showed active coordination behavior: widest gaze dispersion (272~px), largest saccade amplitude (19.9$^\circ$) with the highest saccade velocity (3,886~px/s), the longest mean fixations (497~ms), and the lowest blink rate (9.3/min).
The leader also had the highest gyroscope activity (60.3$^\circ$/s) and widest head-roll range (187$^\circ$), confirming physical scanning of the performance space.
P02 (vocalist) directed 50\% of attention to the leader and virtually none (0.1\%) to the bassist.

\paragraph{Finding 2: Attentional Tunneling with Converging Indicators.}

P04 (bassist) exhibited a cluster of metrics consistent with attentional tunneling: the narrowest gaze dispersion (196~px, 28\% below the leader's), the smallest saccade amplitude (11.6$^\circ$), an elevated blink rate (33.2/min, more than double the 15/min resting baseline), and 66.2\% of all dwell time directed at the leader.
The bassist-leader pair generated 602 bidirectional gaze transitions, forming a tight attentional loop.
This is not merely leader-dependence; the vocalist also focused heavily on the leader (50\%), but the bassist's converging physiological indicators (elevated blink rate, restricted saccades, narrow dispersion) point to stress-driven attentional narrowing rather than strategic monitoring. 
We exclude the drummer's blink rate from tunneling analysis because physical drumming activity inflates the measure.

\paragraph{Finding 3: Asymmetric Dependency and Disconnected Dyads.}

The bassist-leader relationship was asymmetric: P04 directed 66\% of attention to P03, while P03 reciprocated at 50\%, suggesting the leader was aware of the bassist's coordination needs.
Meanwhile, the vocalist and bassist were almost completely disconnected: 0.1\% and 0.3\% mutual dwell time.
The attention matrix revealed a group where coordination flowed almost entirely through the leader, creating both a single point of failure and an uneven cognitive burden.

\begin{table}[t]
    \centering
    \caption{\textit{Needfinding metrics summary.}
    P04 (bassist) shows converging indicators of attentional tunneling; P03 (leader) shows broad coordination scanning. P01 blink rate confounded by physical drumming activity.}
    \label{tab:needfinding}
    \small
    \begin{tabular}{@{}lcccc@{}}
    \toprule
     & \textbf{P01} & \textbf{P02} & \textbf{P03} & \textbf{P04} \\
     & Drummer & Vocalist & Leader & Bassist \\
    \midrule
    Blink rate (/min) & 40.0$^\dagger$ & 14.3 & \textbf{9.3} & \textbf{33.2} \\
    Fix.\ duration (ms) & 393 & 336 & \textbf{497} & 373 \\
    Gaze disp.\ (px) & 253 & 228 & \textbf{272} & \textbf{196} \\
    Saccade ampl.\ ($^\circ$) & 15.5 & 11.9 & \textbf{19.9} & \textbf{11.6} \\
    Leader dwell (\%) & 40.7 & 50.1 & --- & \textbf{66.2} \\
    Trans.\ entropy (bits) & \textbf{3.86} & 3.07 & 3.51 & 3.40 \\
    Gyro activ.\ ($^\circ$/s) & 33.3 & 22.7 & \textbf{60.3} & 51.2 \\
    \bottomrule
    \multicolumn{5}{@{}l@{}}{\textsuperscript{$\dagger$}\scriptsize Confounded by physical drumming activity.}
    \end{tabular}
\end{table}

\paragraph{Implications for the Pipeline.}

These observations constrain each pipeline stage.
At \textbf{Capture}: attention in music practice targets people, not data marks, and those targets move continuously.
Head-mounted eye trackers capture the relevant signals, but AOIs must be dynamically tracked across video frames.
At \textbf{Record}: the same metric (dwell time on the leader) has different meanings depending on role.
Blink rate and gaze dispersion together provide a richer tunneling signal than dwell alone.
At \textbf{Revisualize}: the tunneling target is a person, which makes subtractive intervention socially fraught.
And the disconnected dyads (vocalist-bassist mutual ignorance) represent a coordination gap that no individual mirror system would reveal---only a medium or mediator operating at the group level could surface it.

But this last point is where the needfinding's complexity becomes instructive.
The group-level patterns we observed (the asymmetric dependency loop, the disconnected dyad, the hub-and-spoke topology flowing through the leader) involve interactions among four people with different roles, different needs, and different attentional signatures.
Designing a group-level mediator for this would require simultaneously reasoning about what to diminish, for whom, and when, across all four participants.
That is too many unknowns.

\section{From Group Observation to Individual Intervention}
\label{sec:individual}

The needfinding revealed that the bassist's tunneling and the group's coordination failures are related but separable problems.
The bassist tunnels because of the cognitive demands of playing unfamiliar material while tracking the leader's cues.
The vocalist-bassist disconnection exists because neither has reason, or attentional bandwidth, to monitor the other.
These are different failure modes and likely require different interventions.
The group problem is harder because it involves coordinating between people with different perspectives; the individual problem is hard enough.

Our next step is therefore to focus on individual tunneling reduction during sight-reading.
Sight-reading~\cite{Kopiez2008} is a controlled, measurable instance of the same phenomenon: a musician encounters unfamiliar notation, fixates narrowly on the score, and misses peripheral information such as the instructor's demonstrations, other players' cues, and the broader musical context of what they are reading~\cite{Madell2008}.
Unlike band practice, where targets are other people, sight-reading tunneling is directed at a fixed artifact (the score), which simplifies the Revisualize stage.
You can diminish sheet music without the social complications of diminishing a person.

This scoping reflects a deliberate strategy: isolate the subtractive intervention mechanism in a simpler setting before attempting it in a more complex social one.
Five empirical questions must be answered at the individual level before group-level mediation becomes tractable:

\paragraph{Does Subtractive Intervention break Tunneling at all?}

The feedback loop predicts that diminishing the fixation target will shift gaze, but we do not know whether the shift goes somewhere useful (the instructor, the broader score context) or somewhere arbitrary (the floor, the headset frame).
If diminishment just displaces fixation without improving information uptake, the entire mediator strategy fails regardless of social context.

\paragraph{What is the right Tunneling threshold?}

Our needfinding operationalized tunneling through converging metrics---dwell, dispersion, saccade amplitude, blink rate.
But the thresholds that separate ``focused engagement'' from ``dysfunctional tunneling'' are unknown.
In sight-reading, fixating on the current measure is often appropriate; fixating on the same beat for several seconds while the music moves on is not.
The threshold must be relative to task progression, not absolute.

\paragraph{How should Diminishment ramp?}

Abrupt changes are distracting; overly gradual changes are invisible.
The temporal dynamics of diminishment (onset speed, maximum intensity, and reversal when gaze shifts) need empirical calibration.
Graduated desaturation that ramps over 1--2 seconds and reverses immediately on gaze shift is our starting hypothesis, but we have no data yet.

\paragraph{Does Intervention create dependency?}

If the system trains musicians to rely on external gaze management, their attention allocation may worsen when the system is removed.
Learning transfer, whether musicians who practice with gaze-reactive diminishment develop better unaided scanning habits, is the difference between a crutch and a training tool.

\paragraph{How do individual differences interact with Intervention?}

Our needfinding showed that the drummer's balanced scanning, the vocalist's leader-dependence, and the bassist's tunneling coexisted in the same session. 
Expertise, familiarity with the material, and personality likely all modulate tunneling thresholds.
A system that frustrates experts while under-intervening for novices would be worse than no system at all; our needfinding showed exactly this range of attentional profiles within a single ensemble.
These empirical questions must be answered before attempting the group case.
A group-level mediator that must simultaneously manage multiple people's tunneling, respect their different roles, handle the social meaning of diminishing one person from another's perspective, and avoid cascading diminishment across the ensemble would be building on unknowns.

\section{Discussion}
\label{sec:discussion}

The pipeline, the tensions, and the needfinding data raise the following points of discussion:

\paragraph{Social Targets change the Pipeline's Behavior.}

To our knowledge, most attention-aware systems operate on data marks or environmental features that are static, bounded, and non-social. 
Our needfinding shows that attention in collaborative settings targets people, and this changes every pipeline stage.
Capture must track moving, deformable targets.
Record must disambiguate strategic monitoring from stress-driven tunneling, which requires multimodal evidence (dwell + blink rate + dispersion, as in Table~\ref{tab:needfinding}).
Revisualize must contend with the social meaning of altering how one person perceives another.
The tunneling-target paradox (Tension~3) is specific to social targets and does not arise when the tunneling target is a data mark or a sheet of music.

\paragraph{The Feedback Loop is not Optional.}

Designs that do not account for the loop risk encountering self-reinforcement (Tension~1), the attention paradox (Tension~2), or cascading diminishment (Tension~3) unexpectedly.
Accounting for it means designing with the loop in mind: building in recovery mechanisms that reverse revisualization when gaze shifts, capping the rate at which the scene can change in a single feedback cycle, or using the loop constructively (as AAV's de-emphasis mode does, where the loop drives systematic exploration).
We think the pipeline is a useful tool for other SAXR researchers to stress-test their designs: pick a configuration, trace the loop, and ask whether Stage~3's output creates a problem when it re-enters Stage~1.

\paragraph{Disconnected Dyads need Group-level Capture.}

Our needfinding revealed that the vocalist and bassist were almost completely disconnected (0.1\% mutual dwell).
Neither a mirror (individual feedback) nor a simple medium (shared heatmap) would surface this gap, because it is defined by the \emph{absence} of attention rather than its presence.
Detecting attention that \emph{should be} happening but is not requires a model of expected attention patterns for a given task structure: a mediator that operates on gaps rather than on peaks.
This is ultimately where we want to arrive, but it requires solving the individual tunneling problem first.

\paragraph{Proactive Intervention and the Trust Paradox.}

The pipeline also lets us reason about configurations we have not built.
A proactive subtractive system, one that predicts tunneling before onset and preemptively diminishes, faces a structural problem: successful intervention is invisible.
If tunneling never occurs, the user accumulates no evidence that the system helped. Trust calibration fails because evidence of success is absent.
This is the mirror image of reactive systems, where evidence of failure (the tunneling episode that triggered intervention) is always visible.
Whether users can trust a system whose best outcome is nothing happening is an open empirical question, and one that extends beyond attention-aware systems to any proactive intervention in XR.

\section{Conclusion}
\label{sec:conclusion}

Any attention-aware system that tracks gaze and responds to it follows an Attention-Aware Pipeline (Capture $\to$ Record $\to$ Revisualize).
We have shown that the pipeline's feedback loop, where revisualized content becomes new perceptual input that the system then captures and responds to, generates design tensions whose form depends on how each stage is configured.
Two of these tensions are confirmed by our completed systems: AAV's self-reinforcement and HeedVision's attention paradox.
A third is predicted for a system we have not yet built: the tunneling-target paradox for subtractive intervention.
A needfinding study of four musicians during band practice grounds the trajectory in observed attentional tunneling (66\% leader fixation with converging stress indicators), asymmetric dependency loops (602 bidirectional transitions), and disconnected dyads (0.1\% mutual attention between vocalist and bassist).

These group-level observations motivate our next step, but they also constrain it.
Before we can mediate group attention (coordinating what multiple people see of each other) we need to understand whether subtractive intervention reduces tunneling for a single person reading sheet music.
We bring the pipeline's feedback loop, the needfinding data, and the open empirical questions to the SAXR workshop as tools for reasoning about when, how, and whether to make attention visible in shared immersive environments.

\begin{acknowledgments}
    This work is partially supported by Villum Investigator grant VL-54492 by Villum Fonden and the Aarhus University Research Foundation.
    We also thank VISUS for the ongoing collaboration.
\end{acknowledgments}

\bibliography{references}

\end{document}